\ifx\climdyn\undefined
\documentclass{artikel1}
\usepackage[round]{natbib}

\else
\documentclass[twocolumn,natbib]{svjour3}
\usepackage{mathptmx}
\fi
\usepackage{graphicx}
\newcommand\dg{\ensuremath{{}^\circ}}

\begin{document}
\title{Western Europe is warming much faster than expected}
\author{Geert Jan van Oldenborgh \and 
Sybren Drijfhout \and
Aad van Ulden \and
Reindert Haarsma \and
Andreas Sterl \and
Camiel Severijns \and
Wilco Hazeleger \and
Henk Dijkstra}
\ifx\climdyn\undefined
\else
\journalname{Climate Dynamics}
\authorrunning{van Oldenborgh et al.}
\institute{G. J. van Oldenborgh \and S. S. Drijfhout \and A. van Ulden \and R. Haarsma \and A. Sterl \and C. Severijns \and W. Hazeleger
\at Royal Netherlands Institute of Meteorology\\
P. O. Box 201\\
3730 AE De Bilt\\
Netherlands\\
\email{oldenborgh@knmi.nl}\\ \and 
H. Dijkstra \at Institute for Marine and Atmospheric Research, Utrecht University, Netherlands}
\date{Received: date / Accepted: date}
\fi
\maketitle

\begin{abstract}
The warming trend of the last decades is now so strong that it is discernible in local temperature observations.  This opens the possibility to compare the trend to the warming predicted by comprehensive climate models (GCMs), which up to now could not be verified directly to observations on a local scale, because the signal-to-noise ratio was too low.  The observed temperature trend in western Europe over the last decades appears much stronger than simulated by state-of-the-art GCMs.  The difference is very unlikely due to random fluctuations, either in fast weather processes or in decadal climate fluctuations.  In winter and spring, changes in atmospheric circulation are important; in spring and summer changes in soil moisture and cloud cover.  A misrepresentation of the North Atlantic Current affects trends along the coast.  Many of these processes continue to affect trends in projections for the 21st century.  This implies that climate predictions for western Europe probably underestimate the effects of anthropogenic climate change.

\ifx\climdyn\undefined\else\fi
\end{abstract}

\section{Introduction}

Global warming has been detected in the global mean temperature and on continental\hspace{0pt}-scale regions, and this warming has been attributed to anthropogenic causes \citep{Stott2003,IPCC2007WG1}.  The observed global warming trend agrees well with predictions \citep{Rahmstorf2007}.  However, climate change projections are typically made for much smaller areas.  The Netherlands, for instance, corresponds to a single grid box in most current climate models, but the temperature projections in the KNMI'06 scenarios \citep{KNMIscenarios,KNMIscenariospaper} are based on grid point values of global and regional climate models.  In this region, temperatures simulated by Regional Climate models (RCMs) do not deviate much from GCMs, as the prescribed SST and boundary condition determine the temperature to a large extent \citep{Lenderink2007}.

By now, global warming can be detected even on the grid point scale.  In this paper we investigate the high temperature trends observed in western Europe over the last decades.  First we compare these with the trends expected on the basis of climate model experiments.  These turn out to be incompatible with the observations over large regions of Europe.  The discrepancy is very unlikely due to weather or decadal climate fluctuations \citep{Smith2007,Keenlyside2008}.  Searching for the causes of the unexpectedly fast temperature rise in Europe, we discuss the differences between modelled and observed atmospheric circulation, ocean circulation, soil moisture and radiation, aerosols, and snow cover.

\section{Data}
Many of the results below are obtained in the ESSENCE project, a large ensemble of climate experiments aimed to obtain a good estimate of internal climate variability and extremes \citep{ESSENCE}.  The ESSENCE database contains results of a  17-member ensemble of climate runs using the ECHAM5/MPI-OM climate model \citep{MPI-ECHAM5} of the Max-Planck-Institute for Meteorology in Hamburg.  The version used here is the same used for climate scenario runs in preparation of the IPCC Fourth Assessment Report \citep{IPCC2007WG1}. The ECHAM5 version \citep{ECHAM5_techreport} has a horizontal resolution of T63 and 31 vertical hybrid levels with the top level at 10 hPa. The ocean model MPI-OM \citep{Marsland2003} is a primitive equation $z$-coordinate model. It employs a bipolar orthogonal spherical coordinate system in which the two poles are moved to Greenland and West Antarctica, respectively, to avoid the singularity at the North Pole. The resolution is highest, $\mathcal{O}(\mathrm{20-40\:km})$, in the deep water formation regions of the Labrador, Greenland, and Weddell Seas, and along the equator the meridional resolution is about 0.5\dg. There are 40 vertical layers with thickness ranging from 10~m at the surface to 600~m at the bottom.

The experimental period is 1950-2100. For the historical part of
this period (1950-2000) the concentrations of greenhouse gases (GHG)
and sulphate aerosols are specified from observations, while for the
future part (2001-2100) they follow SRES scenario A1b \citep{SRES}.
The runs are initialised from a long run in which historical GHG
concentrations have been used until 1950. Different ensemble members
are generated by disturbing the initial state of the atmosphere.
Gaussian noise with an amplitude of 0.1~K is added to the initial
temperature field. The initial ocean state is not perturbed.

The findings from the ESSENCE ensemble are backed with results from ensembles from the World Climate Research Programme's (WCRP) Coupled Model Intercomparison Project phase 3 (CMIP3) multi-model dataset.  We used the models with the most realistic circulation selected in \citet{vanUldenvanOldenborgh2006}.  The criterion used was that the explained variance of monthly sea-level pressure fields should be positive for all months.  The explained variance is given by
\begin{equation}
E = 
1 - \frac{\sigma^2_\mathrm{diff}}{\sigma^2_\mathrm{obs}}
\label{eq1}
\end{equation}
Here, $\sigma^2_\mathrm{diff}$ is the spatial variance of the difference between simulated and observed long-term mean pressure, and $\sigma^2_\mathrm{obs}$ the spatial variance of the observed field.  A negative explained variance indicates that the monthly mean sea-level pressure deviates more from the observed field than the reanalysed field deviates from zero.

Apart from ECHAM5/MPI-OM, the models that were selected are the GFDL CM2.1 model \citep{GFDL-CM2.0}, MIROC 3.2 T106 \citep{MIROC3.2}, HadGEM1 \citep{UKMO-HadGEM1} and CCCMA CGCM 3.2 T63 \citep{CGCM3.1}.  Lower-resolution versions of these models also satisfy the criterion, but were thought not to contribute additional information.  Observed greenhouse gas and aerosol concentrations were used up to 2000, afterwards the SRES A1b scenario was prescribed.

The model results are compared with analysed observations in the CRUTEM3 \citep{CRUTEM3} and HadSST2 \citep{HadSST2} datasets.  These have been merged with weighing factors proportional to the fraction of land and sea in the grid box.  For the global mean temperature the variance-weighed HadCRUT3 dataset has been used.  However, this weighing procedure was found to give unrealistic trends in the gridded HadCRUT3 dataset over Europe in summer.  The variance of the HadSST2 grid boxes that are mainly land is very small, so these dominate the combined value, severely down-weighing the CRUTEM3 land observations.

\section{Trend definition}

Trends are computed as the linear regression against the globally averaged temperature anomalies, smoothed with a 3yr running mean to remove the effects of ENSO, over 1950-2007.  This definition is physically better justified than a linear trend \citep[as used in, e.g.,][]{Scherrer2005}, and gives a better signal-to-noise ratio.  In other words, we assume that the local temperature is proportional to the global temperature trend plus random weather noise:
\begin{equation}
T'(x,y,t) = A(x,y) T^{\prime(3)}_\mathrm{global}(t) + \epsilon(x,y,t)\;.
\end{equation}

The difference between observed and modelled trends is described by $z$-values.  These are derived from the regression estimates and their errors:
\begin{equation}
z = \frac{A_\mathrm{obs} - \overline{A}_\mathrm{mod}}{\sqrt{(\Delta A_\mathrm{obs})^2 + (\overline{\Delta A}_\mathrm{mod})^2/N}}
\end{equation}
with $N$ the number of ensemble members and the bar denoting the ensemble average.  The standard errors $\Delta A$ are computed assuming a normal distribution of the trends $A$.  The normal approximation has been verified in the model, where the skewness of the 17 trend estimates is less than 0.2 in almost all areas where $z>2$ in Fig.~\ref{fig:obs_trends}.  Serial correlations have been taken into account whenever significant.

\section{Observed and modelled trends}

\begin{figure*}
\noindent\includegraphics[width=\textwidth]{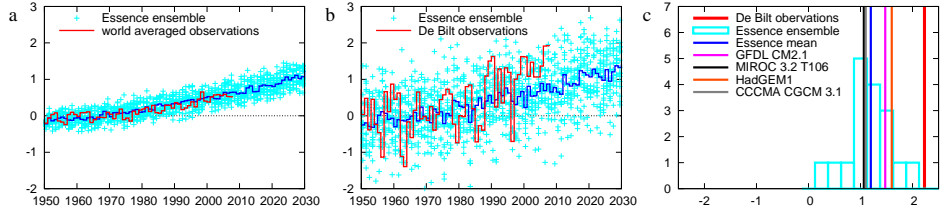}
\caption{Annual mean temperature anomalies [K] relative to 1951-1980 in observations (red) and the ESSENCE ensemble (blue, 17 realisations and the ensemble mean).  a) Global mean, b) De Bilt, the Netherlands (52\dg N, 5\dg E). c) modelled and observed trends [K/K] at De Bilt in the ESSENCE ensemble (histogram) and the four other selected CMIP3 climate models.}
\label{fig:timeseries}
\end{figure*}

Fig.~\ref{fig:timeseries}a shows the global annual mean temperature anomalies from observations (HadCRUT3) and in the 17-member ESSENCE project ensemble.  The model is seen to give a very good description of the warming trend so far; the regression of modelled on observed global mean temperature is $1.06\pm0.06$.

In Fig.~\ref{fig:timeseries}b the temperature at the model grid point representing the Netherlands is compared with observations at De Bilt, corrected for changes in observation practices and warming due to urbanisation \citep{Brandsma2003}.  Random fluctuations due to the weather are much larger at this small spatial scale.  In contrast to the global trends, the local observations show a much stronger warming trend than simulated by this climate model over the last two decades.  The model simulates a factor $1.24\pm0.09$ faster warming than the global mean, but the observations have a trend $A=2.23\pm0.36$.  The histogram of Fig.~\ref{fig:timeseries}c shows that not a single ensemble member has a trend this high over 1950-2007.  The four other selected CMIP3 models also show a trend that is much lower than observed.

\begin{figure}
\noindent\includegraphics[width=\columnwidth]{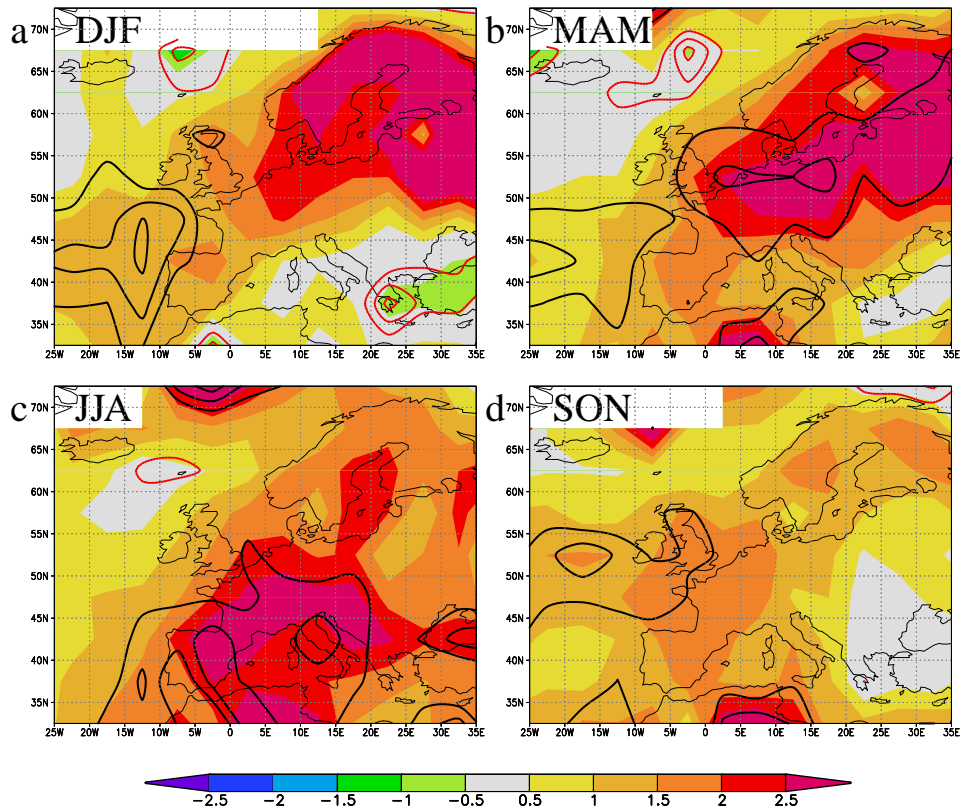}
\caption{Observed trends in surface temperature (colour, [K/K] March 1950 - February 2008, in the merged HadSST2/CRUTEM3 dataset. a) Dec-Feb, b) Mar-May, c) Jun-Aug, d) Sep-Nov.  A value of one denotes a trend equal to global mean warming.  The contours indicate the $z=2, 3$ and $4$ lines of the significance of the difference with the modelled trends (ESSENCE ensemble).  Black (red) indicates that the observed trend is significantly larger (smaller) than the modelled trend.}
\label{fig:obs_trends}
\end{figure}

Maps of the observed warming trends $A(x,y)$ in Europe over 1950 to 2007 are shown in Fig.~\ref{fig:obs_trends}.  As the mechanisms vary over the seasons these are shown separately.  We also show $z$-values for the differences between modelled and observed trends by contours starting at $z=2$.  The areas for which $|z|>2$ correspond to regions where the hypothesis that the model describes the observed trends well can be rejected at the 95\% confidence level.  This area almost coincides with the region where the observed trends are higher or lower than any in the 17-member ensemble.

In all seasons the eastern Atlantic Ocean has warmed significantly faster than the model simulated.  In spring there are also discrepancies of up to 3 standard deviations over land from France to the Baltic and Russia.  In summer, the largest discrepancies are in the Mediterranean area, the $z=2$ contour extending north to the Netherlands.  In autumn, over land only Great Britain has 95\% significant discrepancies between observed and modelled trends.

The area inside the $z=2$ contour, 12\% to 29\% of the area enclosed in 32\dg--72\dg N, 25\dg W--35\dg E, is much larger than the 6\% expected by chance at 95\% confidence. For the $z=3$ contour the area is 2\% to 6\%, larger than the 2.5\% expected except in winter.  The area expected by chance includes the effects of spatial correlations, assuming 30 degrees of freedom \citep{LivezeyChen1983}.

\begin{figure*}
\noindent\includegraphics[width=\textwidth]{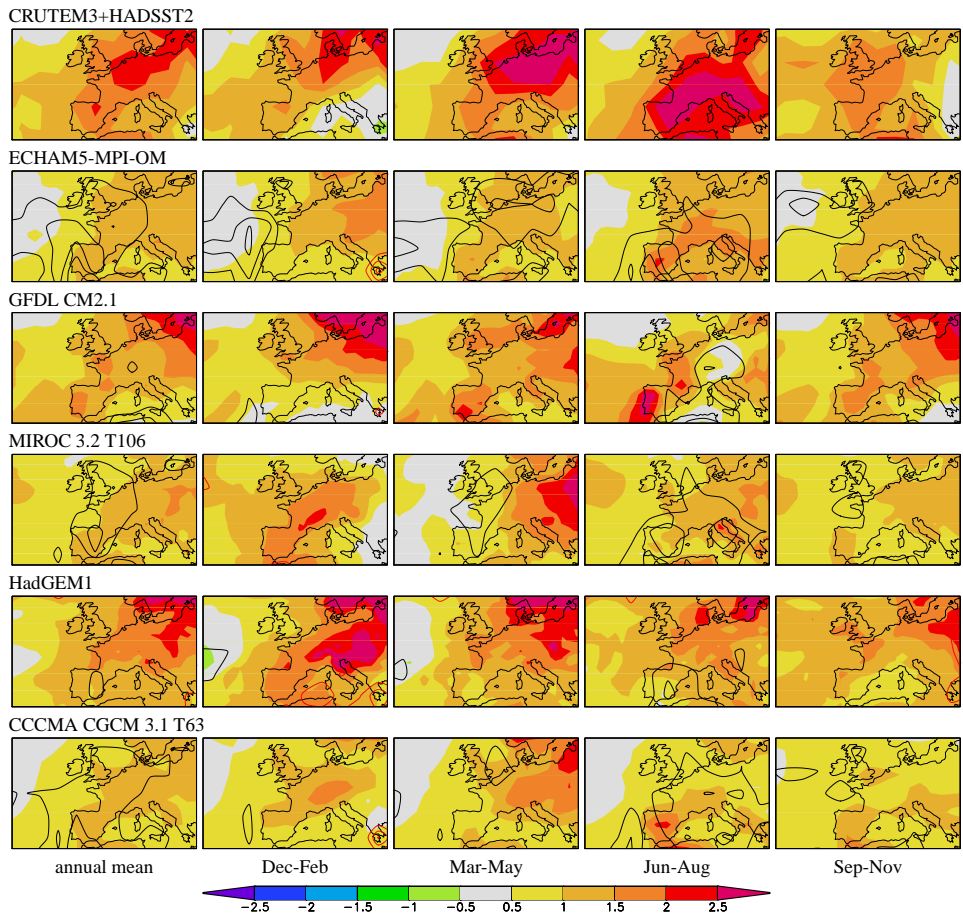}
\caption{The trends in temperature in western Europe as the regression against (modelled) global mean temperature [K/K] in the observations and the GCMs with the most realistic mean circulation in Europe over 1950-2007.  The contours denote the number of standard errors between the observed and modelled trends starting at $z=2$ (black) and $z=-2$ (red).}
\label{fig:trends}
\end{figure*}

We performed similar analyses for four other models used for the IPCC Fourth Assessment Report \citep{IPCC2007WG1} that simulate the current climate in Europe well \citep{vanUldenvanOldenborgh2006}.  In Fig.~\ref{fig:trends} the local temperature trends over 1950-2007 are shown over Europe in the observations, the ESSENCE ensemble of ECHAM5/MPI-OM model runs, GFDL CM2.1, MIROC 3.2 T106, HadGEM1 and CCCMA CGCM 3.2 T63 models.  For the models, we define the trend as the regression against the modelled global mean temperature\footnote{The MIROC 3.2 T106, HadGEM1 and CCCMA CGCM 3.2 T63 experiments in the CMIP3 archive exhibit an $\mathcal{O}(1.5)$ times faster global mean temperature rise than observed.}.  Over western Europe, the patterns of change are similar to the ones in Fig.~\ref{fig:obs_trends}, although the statistical significance is lower due to the smaller ensemble sizes.  Other ensemble simulations appear to exhibit similar behaviour (M. Collins, private communication).  Time slice experiments of the PRUDENCE ensemble of high-resolution regional climate models show temperature changes that are similar to the equivalent GCM changes \citep{ChristensenChristensen2007}.

Figs~\ref{fig:obs_trends},\ref{fig:trends} show that the probability is very low that the discrepancy between observed and modelled warming trends is entirely due to natural variability: the area enclosed by the contours is much larger than expected by chance.  We therefore investigate which physical trends are misrepresented in the GCMs.

\section{Atmospheric circulation}

\begin{figure}
\noindent\includegraphics[width=\columnwidth]{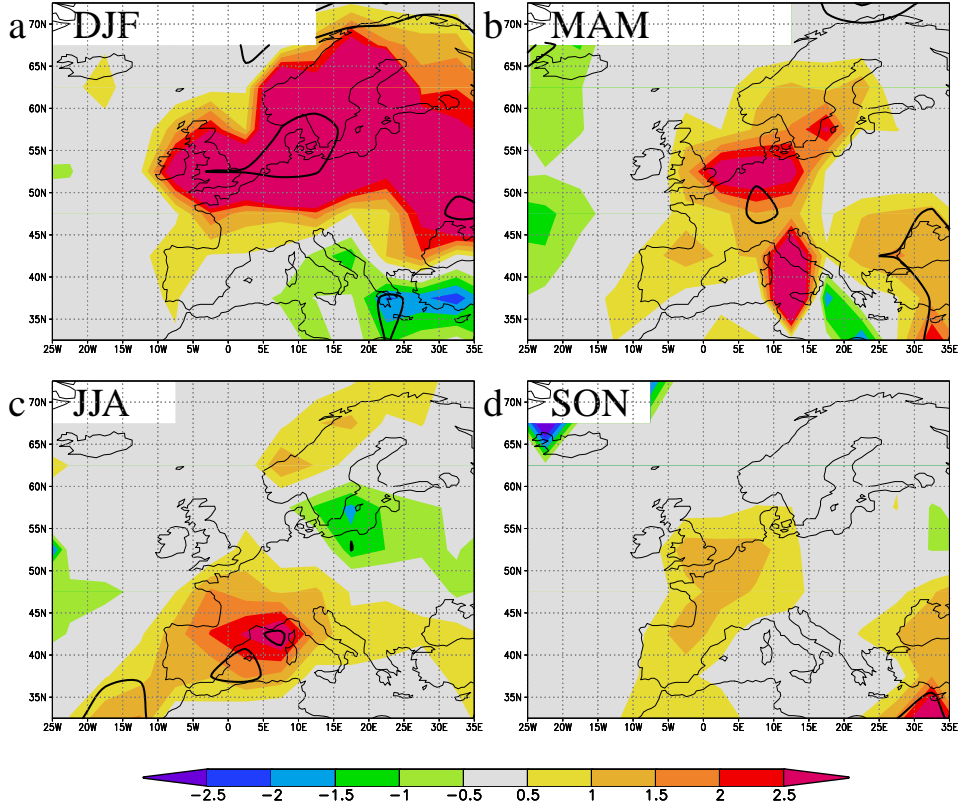}
\caption{As Fig.~\ref{fig:obs_trends}, but for the circulation-dependent temperature $T_\mathrm{circ}$.}
\label{fig:vsm}
\end{figure}

In Europe, at the edge of a continent, changes in temperature are caused to a large extent by changes in atmospheric circulation \citep{OsbornJones2000,Turnpenny2002,vanOldenborghvanUlden2003}.
To investigate the effects of trends in the atmospheric circulation, monthly mean temperature anomalies are approximated by a simple model that isolates the linear effect of circulation anomalies \citep{vanUldenvanOldenborgh2006,vanUlden2007}.  These are the effects of the mean geostrophic wind anomalies $U'(t), V'(t)$ across the temperature gradients, and vorticity anomalies $W'(t)$ that influence cloud cover.  The other terms are the direct effect of global warming, approximated again by a linear dependence on the global mean temperature $T'_\mathrm{global}(t)$, and the remaining noise $\eta(t)$.  A memory term $M$ describes the dependence on the temperature one month earlier, which is important near coasts \citep{vanUldenvanOldenborgh2006}:
\begin{eqnarray}
\label{eq:vsm1}
T'(t) & = & T'_\mathrm{circ} + T'_\mathrm{noncirc}(t) + M T'(t-1)\\
T'_\mathrm{circ} & = & A_U U'(t)
+ A_V V'(t) + B W'(t)\\
T'_\mathrm{noncirc}(t) & = & A T'_\mathrm{global}(t) + \eta(t)\,.
\label{eq:vsm3}
\end{eqnarray}
The geostrophic wind and vorticity anomalies $U',V',W'$ are computed from the NCEP/NCAR reanalysis sea-level pressure \citep{Kalnay1996} and the coefficients $M,A_V,A_U,B$ and $A$ are fitted over 1948-2007 for each calendar month.  This model explains more than half the variance in monthly mean temperature over most of Europe, both in the observations and the models (with coefficients fitted from model data).
Temperature changes that are due to changes in the atmospheric circulation show up as trends in $T'_\mathrm{circ}$.  
Fig.~\ref{fig:vsm} shows the warming trends in the circulation-dependent temperature in the observations and the significance of the difference with the ECHAM5/MPI-OM climate model results.

\begin{figure}
\noindent\includegraphics[width=\columnwidth]{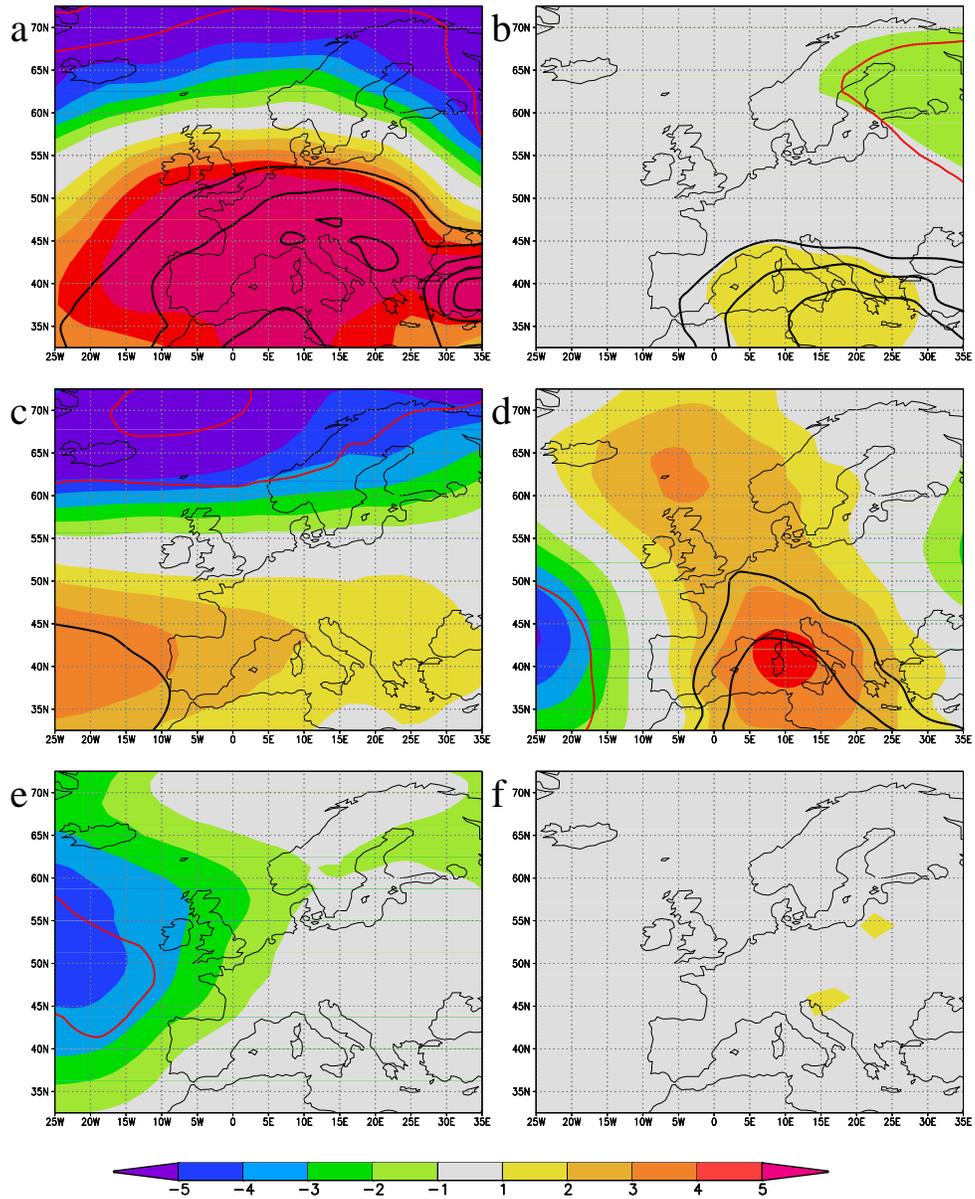}
\caption{Trends in Dec-Feb sea-level pressure [hPa/K] over 1950-2007 in the NCEP/NCAR reanalysis (a), ECHAM5/MPI-OM (b), GFDL CM 2.1 (c), MIROC 3.2 T106 (d), CCCMA CGCM 3.1 T63 (e) and HadGEM1 (f).  The contours denote the $z$-value of the trend being different from zero, starting at 2.}
\label{fig:slp_trends}
\end{figure}

In winter, the observed temperature rise around 52\dg N is dominated by circulation changes.  Fig.~\ref{fig:slp_trends}a shows that a significant increase in air pressure over the Mediterranean \citep{Osborn2004} ($z>3$) and a not statistically significant air pressure decrease over Scandinavia ($z<2$) have brought more mild maritime air into Europe north of the Alps.  

In Fig.~\ref{fig:slp_trends} trends in sea-level pressure over 1950-2007 of the NCEP/NCAR reanalysis are compared to climate model simulations.  Both the reanalysis and the ESSENCE ensemble show a significant trend in the Mediterranean region, but the observed trend is a factor four larger than the modelled trend.  The GFDL CM2.1 and MIROC 3.2 T106 models also show significant positive trends in this area, but again much smaller than observed.  The other two models show no positive trends there.

We conclude that the temperature trends in winter and to a lesser extend spring are due to a shift towards a more westerly circulation.  This change is underrepresented in climate models.  
In summer and autumn the rise in temperature is mainly caused by factors not linearly related to shifts in atmospheric circulation.

\section{Oceanic circulation}

The temperature trend in the  eastern Atlantic Ocean is underestimated by the model results in all seasons but summer and this motivated an investigation of the Atlantic ocean circulation.  The discrepancy may be either a result of ocean memory of the initial state, or model errors.

\begin{figure}
\noindent\centerline{\includegraphics[width=0.7\columnwidth]{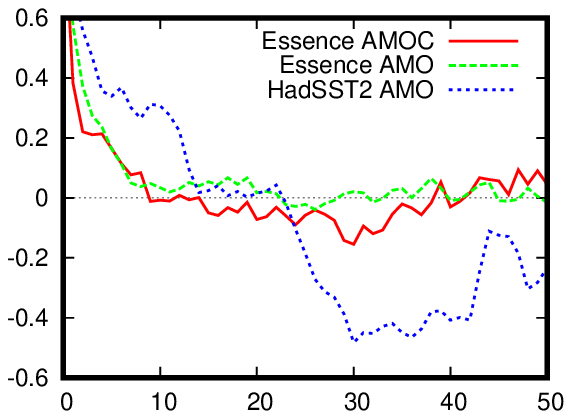}}
\caption{Autocorrelation function of the ECHAM5/MPI-OM Atlantic Meridional Overturning Circulation (AMOC) at 35\dg N and the Atlantic Multidecadal Oscillation (AMO) index, SST averaged over 25\dg--60\dg N, 75\dg--7\dg W.  The effects of external forcing have been minimised by taking anomalies relative to the ensemble mean in the model, and by subtracting the regression against the global mean temperature in the observations.}
\label{fig:amoc}
\end{figure}

The ESSENCE ensemble was started from a common ocean initial state in the model year 1950.  This initial state was taken from a coupled run, so it does not correspond to the real state of the ocean in 1950.  It has recently been shown that ocean memory and dynamics lead to predictability in year 5-10 in the North Atlantic Ocean \citep{Keenlyside2008}.  However, after 10 years the ocean states have decorrelated completely, as is illustrated by the autocorrelation function of the maximum overturning circulation at 35\dg N and an index of the Atlantic Multidecadal Oscillation (AMO) shown in Fig.~\ref{fig:amoc}.  This result is in agreement with the decorrelation time of less than 10 years found in a large ensemble of the CCSM 1.4 model \citep{DrijfhoutHazeleger2007}.  As our definition of the trends does not give weight to temperature variations in the first ten years, when the global mean temperature is almost constant, the effect of ocean memory on the trends is negligible.  The fact that the observed trend is outside the ensemble spread therefore includes the effects of decadal climate variations, to the extent that these are simulated well by the models.

In the observations the multi-decadal oscillations in the Atlantic Ocean are stronger and slower (Fig.~\ref{fig:amoc}) than in the ECHAM5/MPI-OM model.  Over the last decades there has been a rising trend in the AMO index.  To disentangle the effects of the AMO and global warming on temperatures in the North Atlantic region, we subtract a term proportional to the global mean temperature from the SST average, fitted over the 150 years with estimates for both.  In the model, this gives the same result as subtracting the ensemble mean (the AMO has very little effect on the global mean temperature).  Over the relatively short period 1950-2007 we then find virtually no contribution from the AMO on the trend in the observations either.

\begin{figure}
\noindent\includegraphics[width=\columnwidth]{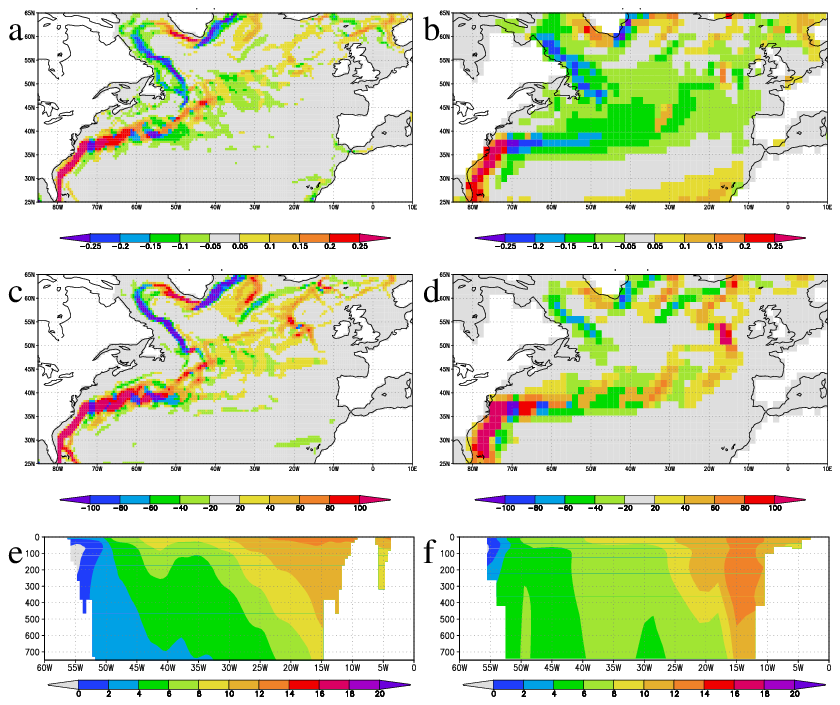}
\caption{Ocean surface currents [$\mathrm{m}\mathrm{s}^{-1}$] in the SODA reanalysis (a) and the ESSENCE ensemble mean (b), both averaged over 1961-1990.  Northward currents are shown positive, southward currents negative, the colour denotes the total velocity. The same for vertically integrated currents from 0 to 750m [$\mathrm{m}^2\mathrm{s}^{-1}$] (c,d).  Subsurface temperature [\dg C] across the Atlantic Ocean at 55\dg N in SODA (e) and the ESSENCE ensemble (f).}
\label{fig:diff_uv}
\end{figure}

Systematic model errors play a much larger role.  The coarse resolution ocean models used in GCMs have a common error in the North Atlantic Current (NAC).  The NAC is compared between the 0.5\dg\ SODA 1.4.1 and 1.4.2 ocean reanalyses \citep{SODA} and the ECHAM5/MPI-OM GCM. Fig.~\ref{fig:diff_uv}c,d show that in the average over the upper 750m, the warm water of the modelled NAC crosses the basin zonally to Portugal, and continues northward, whereas in the reanalysis this Azores current is much weaker and most water meanders north-east across the Atlantic as part of the surface branch of the Atlantic Meridional Overturning Circulation \citep{Lumpkin2003}.

The mean vertical thermal structure is shown in Fig.~\ref{fig:diff_uv}e,f at 55\dg N{}.  The bias in the currents results in a too weak vertical stratification and very deep mixed layers in the modelled East Atlantic, where the surface is cooled by cold fresh water advected from the north (due to too strong westerlies that drive a too large southward Ekman drift, Figs~\ref{fig:diff_uv}a,b) and warmed by the anomalously warm water below (associated with a too far eastward flowing NAC). The deep mixed layer hardly warms under global warming, whereas observed surface temperature rises at about the same rate as the global mean temperature.

\begin{figure}
\noindent\includegraphics[width=\columnwidth]{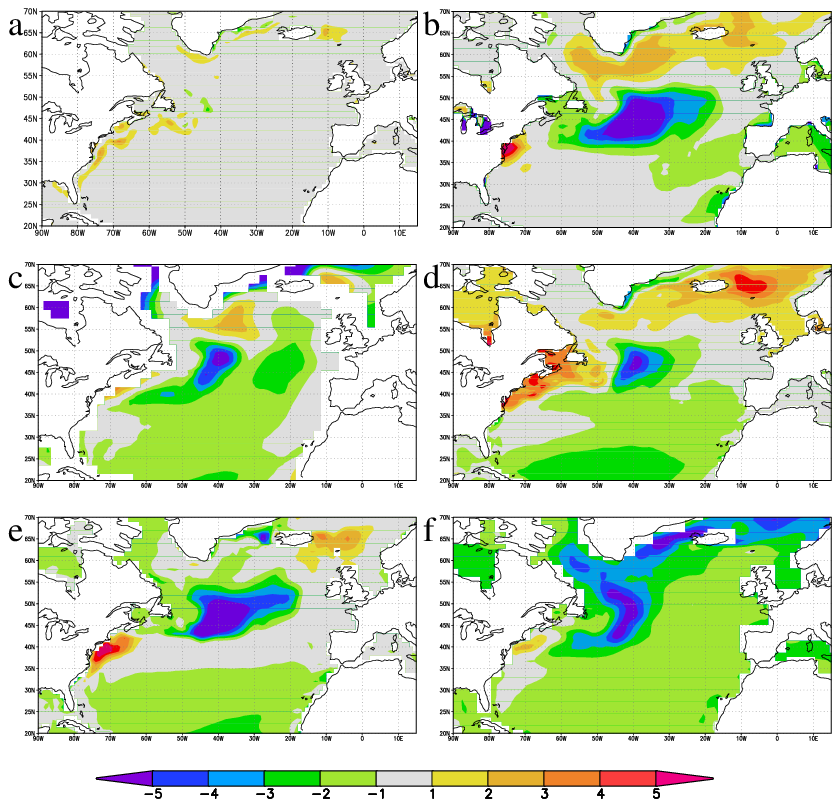}
\caption{Difference between 1982-2007 annual mean SST and the OI v2 SST analysis: SODA ocean reanalysis (a), ESSENCE ensemble (b), GFDL CM 2.1 (c), MIROC 3.2 T106 (d), HadGEM1 (e) and CCCMA CGCM 3.1 T63 (f).}
\label{fig:sst_bias}
\end{figure}

A signature of this bias in the NAC is a strong negative SST bias in the middle of the northern Atlantic Ocean. In the observations this region is south of the NAC, but in the models it is located north of the current and hence it is much colder.  Such a bias is clearly visible in all CMIP3 models considered (Fig.~\ref{fig:sst_bias}b-f), but absent when comparing a the high-resolution SODA reanalysis to the same lower resolution Oi v2 SST analysis \citep[Fig.~\ref{fig:sst_bias}a][]{Reynolds2002}.

\begin{figure}
\noindent\includegraphics[width=\columnwidth]{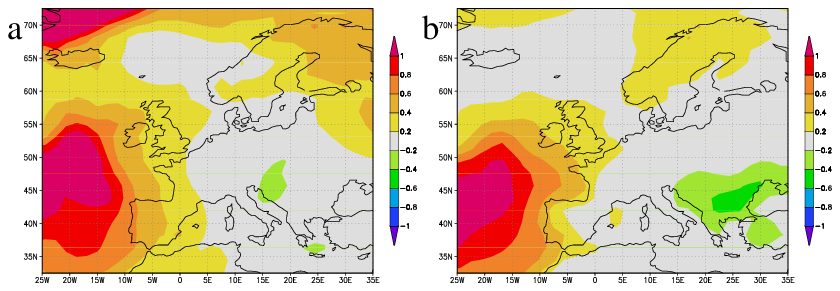}
\caption{Regression of local temperature on SST averaged over 40\dg--50\dg N, 30\dg--10\dg W in the ESSENCE ensemble, low-pass filtered with a 5yr running mean, sum of monthly 1-month lag regressions with SST leading, 1950-2000, anomalies w.r.t.\ the ensemble mean.  Dec-Feb (a), Jun-Aug (b).}
\label{fig:WAtl_to_essence_5yr}
\end{figure}

This bias in the ocean explains the discrepancies over the ocean in Figs~\ref{fig:obs_trends},\ref{fig:trends}.  To estimate the effect on land temperatures, we approximated the effect of a bias in the trend in the East Atlantic on 2-m temperature in Europe by the effect of decadal variability in the same region in the ESSENCE ensemble over 1950-2000.  For each month, the regression of 2-m temperature was computed with SST averaged over 40\dg--50\dg N, 30\dg--10\dg W the previous month, low-pass filtered with a 5yr running mean.  Trends were removed by taking anomalies with respect to the 17-member ensemble mean.  The results are shown in Fig.~\ref{fig:WAtl_to_essence_5yr}.  There is an influence of East Atlantic SST on coastal temperatures of 0.3 to 0.5 K per degree change of East Atlantic SST the previous month, but the signal does not extend very far inland.

\section{Soil moisture and short-wave radiation}

The third important factor explaining discrepancies between observed and modelled trends in Figs~\ref{fig:obs_trends},\ref{fig:trends} consists of related trends in soil moisture and radiation at the surface in spring and summer.  In summer, the pattern of stronger-than-expected heating corresponds closely to the area in which evapotranspiration correlates negatively with temperature in the RCM of \citet{Seneviratne2006} (their Fig.~3a).  This indicates that in this area, the soil moisture is exhausted to the extent that an increase in radiation translates directly into a large increase in temperature, whereas in wetter areas the evapotranspiration increases with rising temperature, damping the high temperatures.  It should be noted, however, that the observed trend ($2.6\pm0.2$ over 40\dg--50\dg N, 0\dg--15\dg E) is much stronger than the modelled trend ($1.4\pm0.1$), indicating that the GCMs underestimate the strength of this process in the current climate.

Regional climate models do not resolve this discrepancy.  Comparing the ES\-SEN\-CE results with the PRUDENCE ensemble \citep{ChristensenChristensen2007}, we find that the second-highest temperature increases in the Mediterranean, the Alps and southern France between 1960-1990 and 2071-2100 are no more than 25\% higher than the equivalent numbers for ECHAM5/MPI-OM, whereas the discrepancy between observed and modelled trends approaches a factor two.  There is therefore no indication that RCMs simulating the last 50 years would show a warming trend as high as observed.

To explain the warming trends further north, we propose a mechanism that closely resembles the mechanism described in \citet{Vautard2007} for extreme summers in Europe.  North of the area with most severe drying, southerly winds bring warmer and drier air northwards, increasing the amount of solar radiation reaching the ground.  Northerly winds do not change.  With the wind direction randomly fluctuating between these two, the net effect is a heating trend accompanied by soil drying.  This way the effects of soil moisture depletion migrate northwards.

We found supporting evidence using Dutch global short-wave radiation observations, which are well-calibrated since the early 1970s \citep{FrantzenRaaff1978}.  The monthly mean observations were corrected for circulation effects using a model analogous to Eqs~(\ref{eq:vsm1})-(\ref{eq:vsm3}).  The trend in circulation is small in late spring and summer (cf.\ Fig.~\ref{fig:vsm}), so subtracting circulation effects mainly decreases the variability.

\begin{figure}
\noindent\includegraphics[width=\columnwidth]{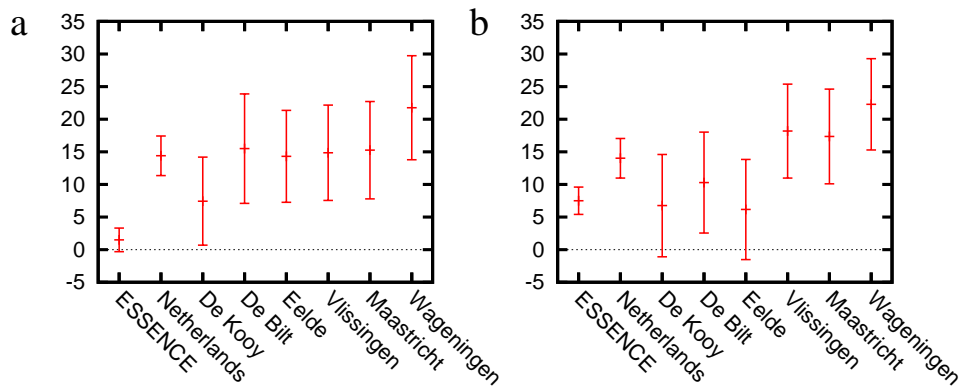}
\caption{Trends over 1971-2007 in global short-wave radiation [$\mathrm{Wm}^{-2}\mathrm{K}^{-1}$] in spring (a) and summer (b) in the ESSENCE ensemble of 17 ECHAM5/MPI-OM model experiments, the Netherlands average, and all stations in the Netherlands.  Error bars denote the standard error.}
\label{fig:qsw_ind_JJA}
\end{figure}

All six stations with observations show an increase in global short-wave radiation in spring and summer over the period 1971-2007, averaging to $14\pm2\:\mathrm{Wm}^{-2}\mathrm{K}^{-1}$ (Fig.~\ref{fig:qsw_ind_JJA}).  To translate changes in short-wave radiation to temperature changes we use a conversion factor obtained from the regression of detrended monthly mean temperature on incoming short-wave radiation, which is $0.05\:\mathrm{K/Wm}^{-2}$.  The observed long-term trend in global short-wave radiation corresponds to roughly 0.7 K warming per degree global mean temperature rise.  This is a sizeable fraction of the total temperature trend, $3.0\pm0.5\:\mathrm{K}/\mathrm{K}$ in spring and $2.2\pm0.6\:\mathrm{K}/\mathrm{K}$ in summer.

The GCM also has a positive trend in this area, but only $5\pm2\:\mathrm{Wm}^{-2}\mathrm{K}^{-1}$ over 1971-2007.  The difference, equivalent to a trend of 0.5 in units of global mean temperature, therefore explains half the discrepancy between observations and model in the Netherlands.  Spatially, the modelled trend in short-wave radiation is at the northern side of the area of strongest warming in Fig.~\ref{fig:obs_trends}c, in accordance with our hypothesis for the summer.  In the model the trend is mainly due to a decrease in cloud cover and continues up to 2100, also supporting the hypothesis that the decrease in cloudiness is driven by soil moisture depletion further south.  We do not have an explanation for the increased sunshine in spring.

There are indications in the observations that the trend is largest on days with southerly wind directions, both in spring and in summer, but the statistical uncertainty on these results is large.  Direct cloud cover observations are unreliable \citep{NorrisWild2007} and uncertainties in cloud cover changes are known to be large \citep{IPCC2007WG1}, making this mechanism difficult to investigate further using observations, but likely to be relevant.

Land use changes are estimated to contribute $\mathcal{O}(0.1\:\mathrm{K})$ to the temperature rise in the Netherlands up to now.  This value comes from a direct estimate of the effect of growing cities around De Bilt \citep{Brandsma2003}.  A rough country-wide estimate can be deduced from the measured increase in `built-up area' of 1\%/10yr over 1986-1996 and 1996-2003 \citep{CBS}.  Assuming the latent heat flux is halved over this area, this decreases evaporative cooling by $\mathcal{O}(2\:\mathrm{Wm}^{-2})$ over 30 years, causing a $\mathcal{O}(0.1\:\mathrm{K})$ temperature rise.  We conclude that land use changes do not contribute substantially to the discrepancy between observed and modelled temperature trends.

\section{Aerosols}

\begin{figure}
\begin{center}
\includegraphics[width=0.7\columnwidth]{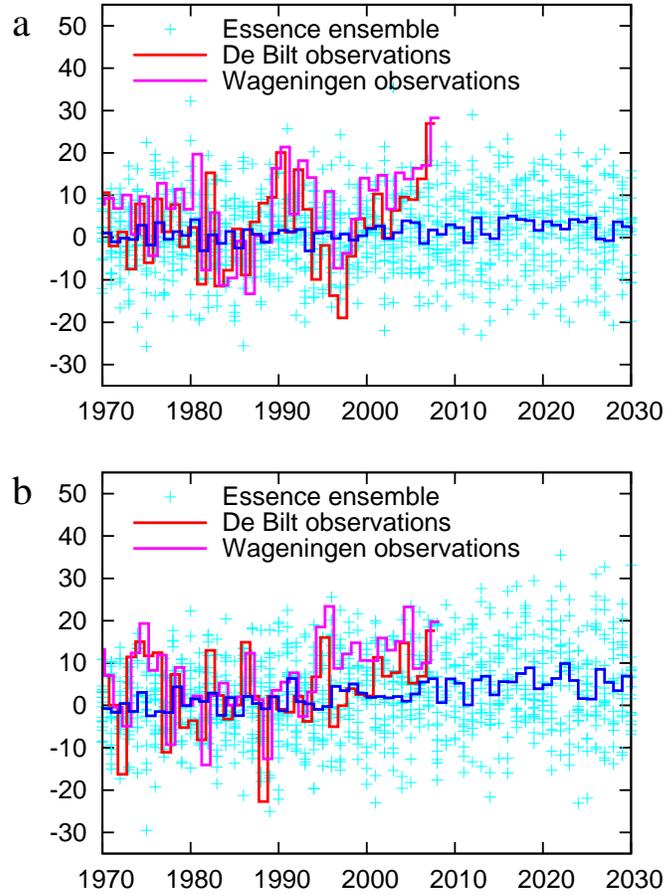}
\end{center}
\caption{Modelled global circulation-independent short-wave radiation [$\mathrm{Wm}^{-2}$] compared with observations at the two stations with the longest records in the Netherlands in spring (a) and summer (b).}
\label{fig:qsw}
\end{figure}

Air pollution has decreased summer temperatures in Europe from 1950 to around 1985, after this clearer skies \citep{Stern2006} have caused a temperature rise \citep{Wild2005,NorrisWild2007,Wild2007}.  This is reflected in first a decrease and later an increase in observed short-wave radiation of about $0.3\:\mathrm{Wm}^{-2}\mathrm{yr}^{-1}$ in the Netherlands in summer (see Fig.~\ref{fig:qsw}).  Converting to an annual mean, this is on the low end of the range quoted European average of $0.3\pm0.1\:\mathrm{Wm}^{-2}\mathrm{yr}^{-1}$, corrected for cloud cover changes \citep{NorrisWild2007}.  As the Netherlands, on the coast, escaped the worst affects of air pollution, this difference is not surprising.

The observed decrease over 1970-1985 translates into a cooling effect of 0.3 to 0.4 K{}.  Note that the effect of this temporary dimming on the trend over the longer period 1971-2007 or 1950-2007 is small: the dimming and brightening cancel each other to a large extend.

In our trend measure the effect of decreased solar radiation due to direct and indirect aerosol effects is about 0.2 times the global mean temperature.  This explains only a small part of the observed trend in the Netherlands in summer.  On shorter time scales, e.g. the period 1985-2007, the reduction of aerosols of course gives a much larger contribution to the temperature trend.

The incoming solar radiation in the ESSENCE ensemble shows a smaller aerosol effect of $0.1\pm0.1\:\mathrm{Wm}^{-2}\mathrm{yr}^{-1}$ in the Netherlands in summer.  The discrepancy translates into a temperature trend bias of only $0.1\pm0.1\:\mathrm{K}$ per degree global warming, significantly smaller than the effect of the bias in long-term trend discussed above.

\section{Snow cover}

\begin{figure}
\noindent\includegraphics[width=\columnwidth]{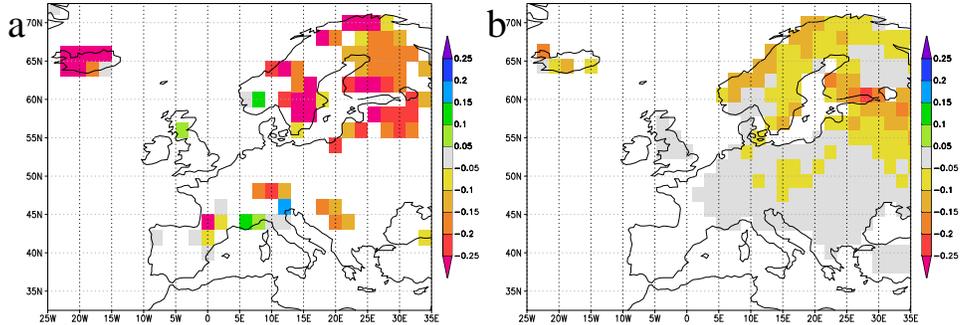}
\caption{Trends in observed (a) and modelled (b) snow cover [$\mathrm{K}^{-1}$] 1972--2007.  Only grid boxes with $p<0.2$ are shown.}
\label{fig:snow}
\end{figure}

In spring, differences in modelled and observed snow cover trends amplify the discrepancies in trends in the Baltic region.  In Fig.~\ref{fig:snow} the trend in Mar-May snow cover is shown in the observations and the ESSENCE ensemble.  The observations indicate a much faster decrease of spring snow cover than the model.  At most grid points the significance of the difference is not very high ($p<0.2$) because of the large decadal fluctuations in the observed snow cover.

\section{Conclusions}

We have shown that the discrepancy between the observed temperature rise in western Europe and the trend simulated in present climate models is very unlikely due to fast weather fluctuations or decadal climate fluctuations.  The main physical mechanisms are varied, both geographically and as a function of the seasonal cycle.  The most important discrepancies between observations and models are
\begin{enumerate}
\item a stronger trend to westerly circulation in later winter and early spring in the observations than in the models,
\item a misrepresentation of the North Atlantic Current in the models giving rise to an underestimation of the trend in coastal areas all year, and
\item in summer, higher observed than modelled trends in areas in southern Europe where soil moisture depletion is important, and consequently a stronger trend in sunshine around the Netherlands in spring and summer.
\end{enumerate}
Smaller contributions come from differences between observed and modelled trends in aerosol effects in spring and summer, and snow cover changes in the Baltic in spring.

As most projections of temperature changes in Europe over the next century are based on GCMs and RCMs with the biases discussed above, these projections are probably biased low.  To correct the biases, it is essential to not only validate the GCMs for a good representation of the mean climate, but also on the observed temperature trends at regional scales.

\ifx\climdyn\undefined
\paragraph{Acknowledgements}
\else
\begin{acknowledgements}
\fi
The ESSENCE project, lead by Wilco Hazeleger (KNMI) and Henk Dijkstra (UU/IMAU), was carried out with support of DEISA, HLRS, SARA and NCF (through NCF projects NRG-2006.06, CAVE-06-023 and SG-06-267). We thank the DEISA Consortium (co-funded by the EU, FP6 projects 508830 / 031513) for support within the DEISA Extreme Computing Initiative (www.deisa.org). The authors thank HLRS and SARA staff for technical support.

We acknowledge the other modelling groups for making their simulations available for analysis, the Program for Climate Model Diagnosis and Intercomparison (PCMDI) for collecting and archiving the CMIP3 model output, and the WCRP's Working Group on Coupled Modelling (WGCM) for organising the model data analysis activity.  The WCRP CMIP3 multi-model dataset is supported by the Office of Science, U.S. Department of Energy.

David Stephenson is thanked for helpful suggestions for the statistical analysis.
\ifx\climdyn\undefined
\bibliographystyle{plainnat}
\else
\end{acknowledgements}
\bibliographystyle{spbasic}
\fi


\end{document}